%% file: N1365_accepted.tex
\documentclass[traditabstract]{aa}
\usepackage{graphics}
\usepackage{amssymb, amsmath}
\usepackage{natbib}
\usepackage{rotating}
\bibpunct{(}{)}{;}{a}{}{,}
\input{thesiscom.tex}
\title{Cold dust in giant barred galaxy NGC~1365}

\author{ F.\,S. Tabatabaei\inst{1}, A. Wei\ss \inst{2},  F. Combes\inst{3}, C.~Henkel\inst{2,4},  K.\,M.~Menten\inst{2}, R. Beck\inst{2},  A. Kov\'acs\inst{5,6}, R. G\"usten\inst{2}}
\authorrunning{Tabatabaei et al.}

\institute{Max-Planck-Institut f\"ur Astronomie, K\"onigstuhl 17, 69117 Heidelberg, Germany
\and Max-Planck Institut f\"ur Radioastronomie, Auf dem H\"ugel 69, 53121 Bonn, Germany 
\and Observatoire de Paris, LERMA, CNRS, 61 Av. de l'Observatoire, 75014 Paris, France
\and Astron. Dept., King Abdulaziz University, P.O. Box 80203, Jeddah, Saudi Arabia
\and University of Minnesota, 116 Church St SE, Minneapolis, MN 55414, USA
\and California Institute of Technology 301-17, 1200 E. California Blvd, Pasadena, CA 91125, USA
}
\offprints{F.\,S. Tabatabaei\\ taba@mpia.de}

\begin{document}

\titlerunning{Cold dust in NGC\,1365}
\authorrunning{Tabatabaei et al.}
\abstract
{Observations of galaxies at sub-millimeter wavelengths, where the emission is mainly due to cold dust, are required to constrain the dust physical properties and provide important insight on the gas content of galaxies. We mapped NGC\,1365 at 870\,$\mu$m with LABOCA, the Large APEX Bolometer Camera, allowing us to probe the central mass concentration as well as the rate at which the gas flows to the center.
We obtained the dust physical properties both globally and locally for different locations in the galaxy. 
A 20\,K modified black body represents about 98\% of the total dust content of the galaxy, the rest can be represented by a warmer dust component of 40\,K.  The bar exhibits an east-west asymmetry in the dust distribution: The eastern bar is heavier than the western bar by more than a factor of 4. Integrating the dust SED, we derive a total infrared (IR) luminosity, L$_{\rm TIR}$, of $9.8 \times\,10^{10}\,{\rm L}_{\odot}$ leading to a dust-enshrouded star formation rate of ${\rm SFR}_{\rm TIR}\simeq$\,16.7\,M$_{\odot}$\,yr$^{-1}$ in NGC\,1365.  We derive the gas mass from the measurements of the dust emission leading to a CO-to-H$_2$ conversion factor of $X_{\rm CO}\simeq 1.2\times 10^{20}$\,mol\,cm$^{-2}$\,(K\,km\,s$^{-1}$)$^{-1}$ in the central disk including the bar.  Taking into account the metallicity variation, the central gas mass concentration is only  $\simeq$\,20\% at $R<40\arcsec$\,(3.6\,kpc).  On the other hand, the time-scale with  which the gas flows into the center, $\simeq$\,300\,Myr, is rather short. This indicates that the current central mass in NGC\,1365 is evolving fast due to the strong bar. 
}
\maketitle

%
%
\section{Introduction}
\label{sec:intro}
Bars are generally considered as an important transform mechanism of molecular gas towards the central regions of galaxies, fueling  central starbursts and active nuclei. This is confirmed by an enhancement of CO emission along the bar \citep[e.g.][]{Gerin,Benedict,Sakamoto_999} and by the resolved offset ridges along the leading edges of the rotating bar \citep{Ishizuki}. However, key questions about the formation and evolution of bars and the influence of bars on the physical and chemical evolution of the interstellar medium  remain open. 
Numerical simulations suggest that barred galaxies tend to have more of their gas mass concentrated in their centers  than non-barred galaxies \citep[e.g.][]{Combes_85}. This is tentatively confirmed by $^{12}$CO(J=1-0) observations \citep{Sakamoto_99}. However, the observational evidence to date is sparse, and has its shortcomings.
On the other hand, it is possible that the central mass concentration is affected by more than just the presence of bars. For example, \cite{Komugi} show that the Hubble type could play a more important role than bars. 
As such, the distribution of gas contained in the disk of barred galaxies could shed light on the issue. 

Although CO observations directly probe the gas in its molecular phase (H$_2$), there are indications that bars can contain gravitationally-unbound molecular gas \citep[e.g.][]{Das,Hutt}. Thus, in barred environments the mass of the molecular gas might be overestimated since the standard Galactic CO to H2  mass factor XCO  (the Virial X-conversion factor) is not necessarily applicable. Moreover, the dependence of the X$_{\rm CO}$ conversion factor on the metallicity \citep{Wilson_95} and the optical thickness of the CO line introduce an uncertainty on the estimate of the total gas mass, generally.

As an alternative method to measure the gas mass, observations of the dust continuum
emission have been suggested and used by several authors \citep[e.g.][]{Hildebrand,Guelin_93,James_02}.
Furthermore, studies based on the $\gamma$-ray observations of the Milky Way with the EGRET \citep{Grenier} and Fermi \citep{Abdo_10_1} space telescopes indicate that dust is a promising tracer of the gas,  even of gas invisible in HI and CO (the so called `dark gas'). 
Detailed studies of the mm and sub-mm continuum emission from the Milky Way and other nearby galaxies show that about 90\% of the dust mass is as cold as 14--16\,K, and that dust is well-mixed with molecular gas so that  cold dust emission can be used to probe the molecular hydrogen \citep[e.g.][]{Misiriotis_06}. The cold dust can be best studied at sub-mm wavelengths. Moreover, the importance of the sub-mm data to constrain the dust spectral energy distribution (SED) and extract dust mass and temperature is already indicated by numerous studies \citep[e.g.][]{Gordon_10}.  Therefore, sub-mm observations of barred galaxies are important to study the physics of the dominant component of the interstellar medium in both disks and bars. 

The total infrared emission (integrated in the wavelength range from, e.g., 8 to 1000\,$\mu$m) is known to be a good tracer of the embedded star formation in galaxies \citep[see][and references therein]{Kennicutt_12}. Nevertheless, it is  still not clear how much of this emission is linked to dust heating sources other than the ongoing star formation e.g. to non-ionizing UV photons or old stellar population.

We investigate the central mass concentration and the physical properties of the cold dust in the ``Great Barred Spiral Galaxy'' NGC~1365.   With a diameter of twice the Milky Way ($\sim$\,60\,kpc) and a mild inclination ($\sim$\,$41^{\circ}$),  NGC~1365 
is among the best studied barred galaxies from the X-ray to the radio regimes, providing a rich multi-wavelength data archive ideal for in-depth studies. This galaxy hosts a Seyfert 1.5 AGN \citep{Schulz} as well as strong star formation activity (starburst) in the center  \citep[e.g. see][and references therein]{Lindblad_99}. NGC~1365 has a nuclear bar of about one kpc embedded in the  large-scale bar \citep{Jungwiert}. This galaxy does not host a circumnuclear ring unlike  many barred galaxies. The shape of the central starburst region is
asymmetric, with two massive dust lanes, with strong and aligned
magnetic fields  \citep{Beck_etal_05}. NGC\,1365 has been observed with the 1.8-m Ballon-borne Large Aperture Submm Telescope (BLAST) at 250, 350, and 500$\mu$m  at resolutions 36\arcsec, 42\arcsec, and 60\arcsec, respectively \citep{Wiebe}. At these wavelengths, the central part of the galaxy has also been observed with the Herschel Space Observatory \citep{Alonso}. 
Here, we present submm observations of this galaxy at 870$\mu$m with the APEX bolometer camera (LABOCA) at a resolution of about 20\arcsec, which is much better than that of the  BLAST and Herschel submm data at 500\,$\mu$m. 

Through a comparison with various tracers of the interstellar medium (ISM), we study the energy sources of the 870\,$\mu$m emission. We also  re-visit the dust physical properties like temperature, mass, and total infra-red luminosity $L_{\rm TIR}$ and use this information to estimate the X$_{\rm CO}$ conversion factor as well as star formation rate in NGC\,1365. In a different approach, we also present the dust physical properties along the bar using the LABOCA 870$\mu$m and the BLAST 250$\mu$m data in apertures of 36\arcsec. The 870\,$\mu$m data is further used as a constraint for a gas flow model in this barred system.  

The paper is organized as follows. The 870\,$\mu$m observations and data reduction as well as the relevant auxiliary data sets used are described in Sect. 2. We investigate the morphology and origin of the 870\,$\mu$m emission and derive the dust physical parameters in Sect.~3. Based on these results, the gas mass concentration and the role of $X_{\rm CO}$ conversion factor are discussed in Sect.~4. We also update estimates of the star formation rate as well as the rate of the gas flow in the center.  The final results are then summarized in Sect.~5.  
\begin{table}
\begin{center}
\caption{Positional data adopted for NGC~1365.}
\begin{tabular}{ l l } 
\hline
\hline
Position of nucleus$^{1}$    & RA\,=\,$03^{h}33^{m}36.37^{s}$      \\
  \,\,\,(J2000)  &  DEC\,=\,$-36^{\circ}08\arcmin25.4\arcsec$\\
PA line of nodes $^{2}$   & 220$^{\circ}$ \\
Inclination$^{2}$   & 41$^{\circ}$  (0$^{\circ}$=face on)\\
Distance$^{3}$   & 18.6\,Mpc\\
\hline
\noalign {\medskip}
\multicolumn{2}{l}{$^{1}$ \cite{Lindblad_96}}\\
\multicolumn{2}{l}{$^{2}$ \cite{Jorsater_Moorsel} }\\
\multicolumn{2}{l}{$^{3}$ \cite{Madore_99}, 1$\arcsec$=\,90.2\,pc }\\
\end{tabular}
\end{center}
\end{table}
\begin{table*}
\begin{center}
\caption{NGC~1365 data used in this study. }
\begin{tabular}{ l l l} 
\hline
Wavelength & Telescope  &   \,\,\,\,\,\,\,Reference \\
\hline
870\,$\mu$m             &   APEX & This paper\\
250-500\,$\mu$m               &   BLAST &\cite{Wiebe}\\
2.6\,mm\,$^{12}$CO(1-0)           &  SEST & \cite{Sandqvist_95} \\
867\,$\mu$m\,$^{12}$CO(3-2)           &  SEST & \cite{Sandqvist_99} \\
21\,cm\,HI               &   VLA &\cite{Jorsater_Moorsel}\\
6.2\,cm                    &   VLA & \cite{Beck_etal_05}\\
1.5\,$\mu$m               & 2MASS & \cite{Jarrett}\\
120-200\,$\mu$m          & ISO-ISOPHOT & \cite{Spinoglio}\\
43-197\,$\mu$m          & ISO-LWS & \cite{Brauher}\\
1516\AA{}\,(FUV)           &   GALEX& \cite{GildePaz_07}\\
\hline
\end{tabular}
\end{center}
\end{table*}
\begin{figure}
\resizebox{7cm}{!}{\includegraphics*{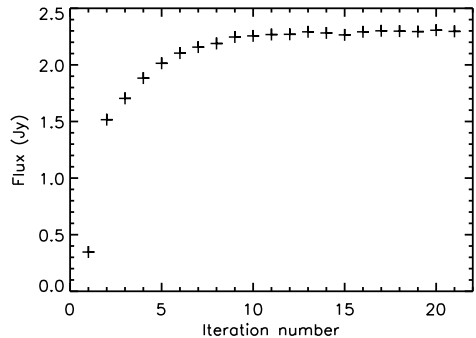}}
\caption[]{Integrated flux density obtained after each iteration in the data reduction which shows a convergence after the 10th iteration.   }
\label{fig:iter}
\end{figure}
\begin{figure*}
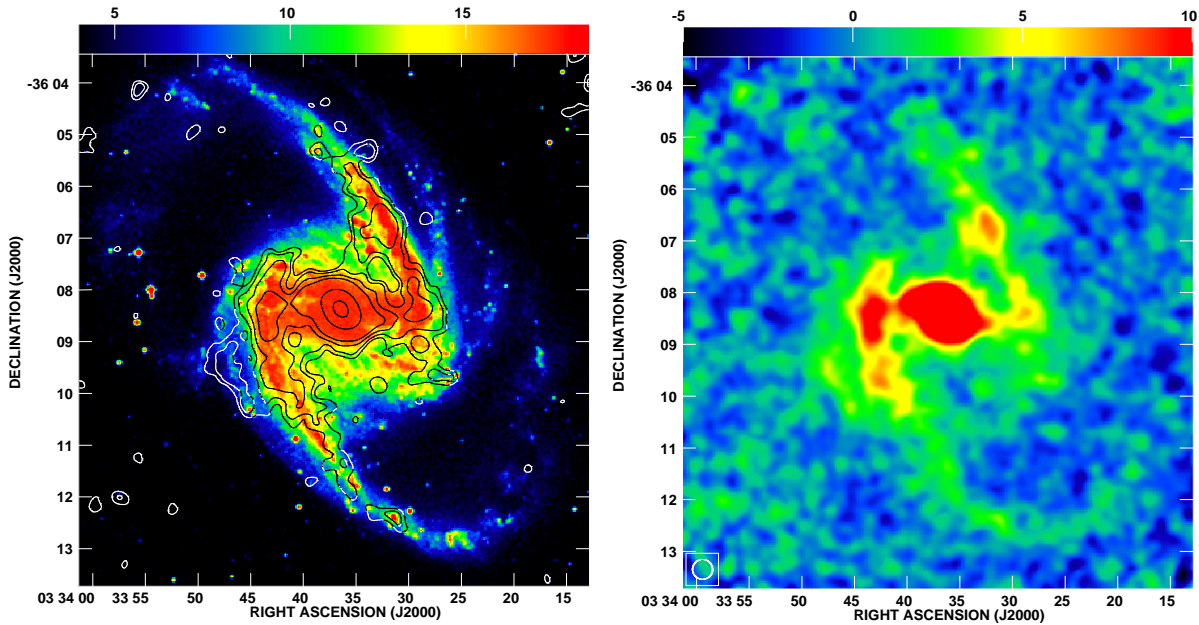

\resizebox{\hsize}{!}{\includegraphics*{N1365-Blue-870cont1.ps}\includegraphics*{n1365-LABOCA-BOA-signaltonoise.ps}}
\caption[]{{\it Left}: submm 870\,$\mu$m emission (contours) superimposed on an optical image (B-band, taken from the STScI Digitized Sky Survey) of NGC\,1365. The contour levels are 6, 9, 15, 24, 150, 500\,mJy/beam. The bar shows the optical surface brightness in arbitrary units. {\it Right}: submm 870\,$\mu$m emission, normalized to the one $\sigma $ noise rms level (signal-to-noise ratio).  The angular resolution of 23$\arcsec$ is shown in the lower left corner.}
\label{fig:submm}
\end{figure*}
\begin{figure*}
\resizebox{7.8cm}{!}{\includegraphics*{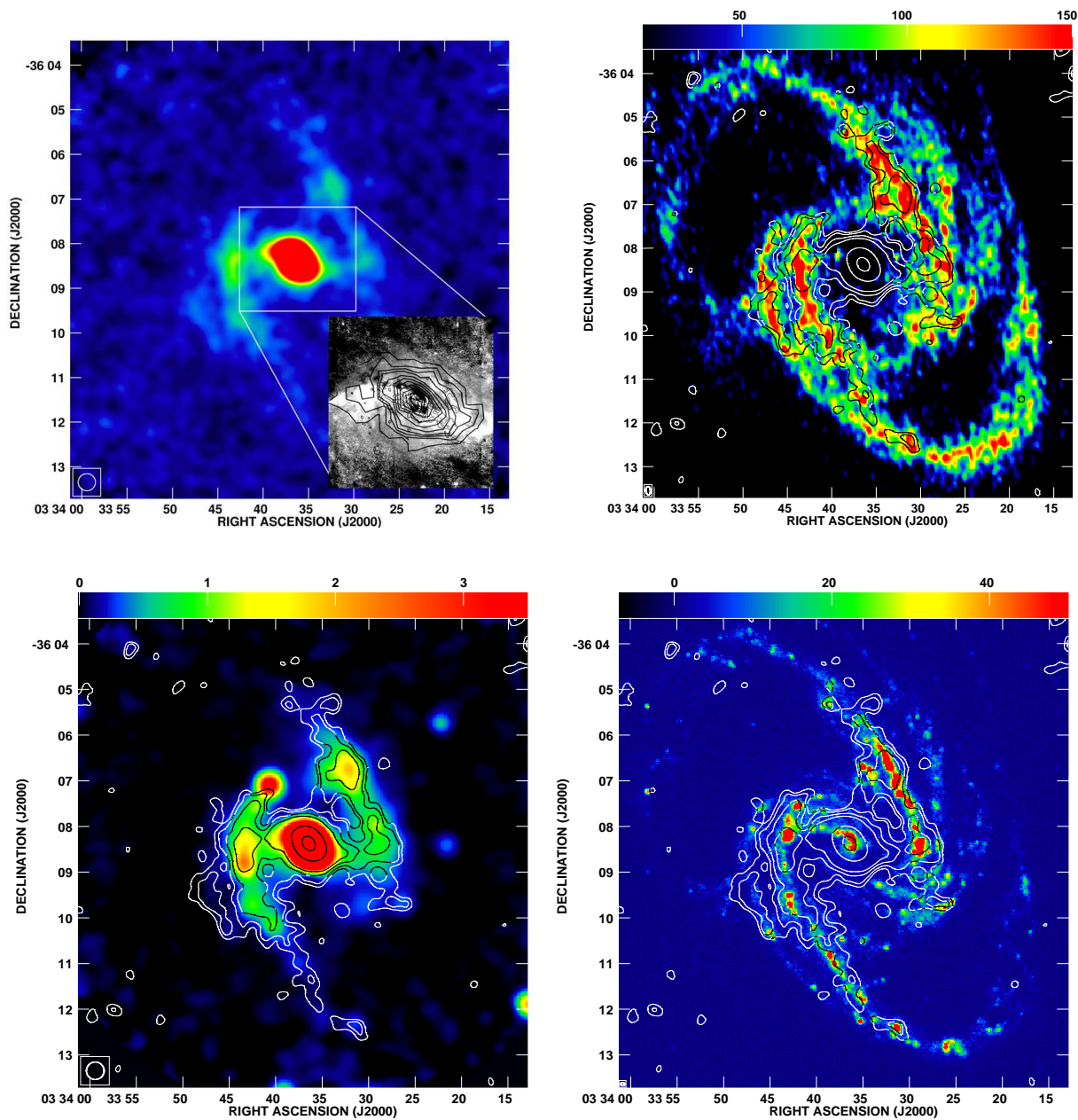}}
\resizebox{7.4cm}{!}{\includegraphics*{N1365-870um.Boa_new.23arc.contour+HI_3.ps}}
\resizebox{\hsize}{!}{\includegraphics*{n1365-LABOCA-BOAnew-6cm_3.ps}
\includegraphics*{n1365-LABOCA+FUV_3.ps}}
\caption[]{ {\it Top left:} LABOCA 870\,$\mu$m emission compared to  CO(2-1) observations of the  central part of NGC~1365 \citep[contours on top of an optical image, see ][]{Sandqvist_95}.  {\it Top right:} Contours of the 870\,$\mu$m emission overlaid on the HI map.  {\it Bottom left:} The same contours on top of the radio continuum emission at 6\,cm and the GALEX FUV map ({\it bottom right}). In all panels the  
resolution of the 870\,$\mu$m emission is 23$\arcsec$ with contour levels of 6, 9, 15, 24, 150, 500\,mJy/beam.   }
\label{fig:compare}
\end{figure*}
\section{Data}
\subsection{Sub-mm observations and data reduction}
The 870\,$\mu$m data were taken with the Large APEX BOlometer Camera \citep[LABOCA][]{Siringo}, a 295-pixel bolometer array, operated on the Atacama Pathfinder EXperiment 12 meter diameter telescope \citep{Guesten} in Chanjantor, Chile. We observed NGC~1365 in 2008  December and 2009 August in mostly good weather conditions (the precipitable water vapour PWV content ranged from 0.1\,mm to 0.9\,mm). NGC~1365 was mapped in the spiral raster mode providing a fully sampled map in the LABOCA FOV ($11\arcmin \times 11\arcmin$) in each scan. The total on-source integration time was about 12 hours. The data were calibrated by observing Mars and Uranus together with the secondary calibrators and was found to be accurate within 15\%. 
The data were reduced using the BOA (BOlometer array Analysis)
software  \citep{Siringo,Schuller}. After flagging for bad and noisy pixels, the data were despiked and correlated noise was removed for each scan. 
Then the scans were coadded (weighted by rms$^{-2}$) to create the final map. 

This process was performed 21 times in an iterative approach following \cite{Belloche}. After a first iteration of the reduction, we made a source model by setting the map  to zero below a signal-to noise ratio of 4. Then the source map was used to flag bright sources and the data were reduced again. After the fourth iteration, the map resulting from the previous iteration was set to zero below a signal-to-noise ratio of 2.5. The remaining signal was subtracted from the data before reduction and added back after reduction. 
This way,  negative artifacts which appear around the bright sources are much reduced, more extended emission can be recovered,  and a more stable background noise level in the central region is obtained. Figure~\ref{fig:iter} shows a fast increase in the integrated flux density  from the first to the 5th iteration, reaching to a stable situation after the 10th iteration.

The HPBW of the telescope at 870\,$\mu$m is 19.2$\arcsec$. The map was convolved to 23$\arcsec$ in order to achieve a better signal-to-noise ratio without loosing too much spatial information about the emission properties. Figure~\ref{fig:submm} shows the convolved map with an rms noise of 3\,mJy/beam. 
\subsection{Complementary data}
This study is supplemented  with other tracers of the neutral and ionized gas.    Table 2 summarizes the data used in this work.  \cite{Jorsater_Moorsel} mapped NGC~1365 in 21-cm HI line emission with the VLA using hybrid BnA, CnB, and DnC configurations at a resolution of $11.6\arcsec \times 6.3\arcsec$ (Fig.~\ref{fig:compare}). This dataset has been corrected for missing spacings.  NGC~1365 was observed in $^{12}$CO(1-0)  over a $204\arcsec \times 164\arcsec$ region centered on the nucleus with the Swedish/ESO Submillimeter Telescope (SEST) by \cite{Sandqvist_95} at a resolution of 44\arcsec. In order to subtract its contribution in the LABOCA band, the SEST observations of the $^{12}$CO(3-2) line  \citep{Sandqvist_99}  are used as well. 

\cite{Wiebe} presented the BLAST observations of NGC~1365 at 250, 350, and 500\,$\mu$m  at resolutions 36$\arcsec$, 42$\arcsec$, and 60$\arcsec$, respectively. We used their maps clipped in an area of $13\arcmin \times 13\arcmin$ centered on the nucleus. Moreover, the FIR measurements of ISOPHOT \citep{Spinoglio}, ISO-LWS \citep{Brauher} made with the Infrared Space Observatory (ISO) as well as with the Infrared Astronomical Satellite (IRAS) \citep{Sanders_03} have been used to study the dust SED.  

The radio continuum emission from NGC~1365  was mapped with VLA at 6.2\,cm and at 13\arcsec resolution \citep{Beck_etal_05}. The radio 6.2\,cm emission is mainly emerging from the central $300\arcsec \times 300\arcsec$ region. We used the 6.2\,cm map after subtracting the bright background radio source in the north-east of the galaxy. 
In far ultraviolet (FUV),  NGC~1365 was observed with the GALaxy Evolution EXplorer satellite (GALEX) at 4.5\arcsec resolution as detailed in the GALEX ultraviolet atlas of nearby galaxies \citep{GildePaz_07}.
\section{Results}
\subsection{Morphology of the 870\,$\mu$m continuum map}
NGC~1365 is illuminated by its oval-shape core of $\sim$\,80\arcsec diameter at 870\,$\mu$m (Fig.~\ref{fig:submm}). In this region, the 870\,$\mu$m intensities are higher than 100\,mJy/beam with a maximum of $\sim$\,600\,mJy/beam. The bar is brighter in the eastern edge than in the western edge. The  two main spiral arms appear pronounced by bright clumps corresponding to the complexes of star forming regions followed by faint emission ($\sim 3\sigma$) in the outer parts. A segment of the secondary arm in the south-east of the galaxy, which is weak in optical images but bright in HI, is detected at 870\,$\mu$m as well (see the 870\,$\mu$m contours overlaid on a HI map in Fig.~\ref{fig:compare}). 
Apart from their similarity along the spiral arms, the 870\,$\mu$m and the HI emission show a striking difference in the central part including the nucleus and the bar: While this region is the brightest part at 870\,$\mu$m, it is the darkest in HI.  The central part is also the most dominant region in the CO(1-0) line emission as well as in the radio continuum emission (e.g. at 6\,cm, Fig.~\ref{fig:compare}).  While \citep[e.g. ][]{Ondrechen} find weak HI absorption in a limited velocity range, the virtually complete absence of HI toward the  center of NGC\,1365 and most of its bar is explained by the fact that almost all the gas in these dense regions is in molecular form (and traced by the strong CO emission). 
Generally, the 870$\mu$m emission is very similar to the 6\,cm radio continuum emission, particularly, in the extent of the main arms and the east-west asymmetry of the bar. On the other hand, the radio continuum emission is very weak in the  secondary arm in the south-east of the galaxy which is bright at 870$\mu$m (and HI).  This must be a region of high gas density, but with only little star formation. In the 6\,cm radio continuum map, the strong source in the north-east is a background radio source \citep{Sandqvist_82}. In the FUV, the core does not dominate the emission. The strong central 870$\mu$m emission indicates a significant attenuation of the UV emission by  dust emitting in the FIR/submm range. 

The integrated flux density of the 870$\mu$m emission in the plane of the galaxy (using parameters listed in Table~1) around the center out to a radius of 220\arcsec (20\,kpc) is $S\,=\,2.3 \pm 0.3$ Jy. 
The integrated flux density in the core ($R<40\arcsec$ or 3.6\,kpc) is $S\,=\,1.1 \pm 0.2$ Jy, about half of the total value.
\begin{figure}
\resizebox{\hsize}{!}{\includegraphics*{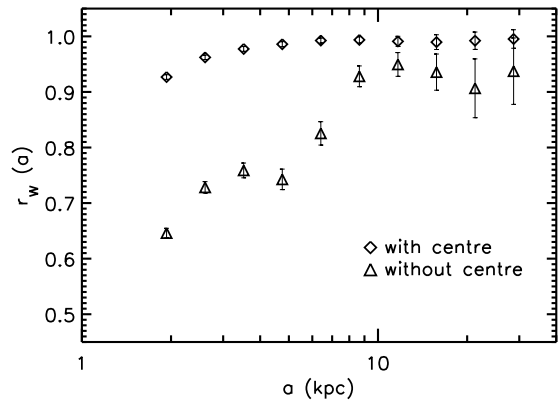}}
\caption[]{The scale-by-scale correlation between the cold dust emission at 870\,$\mu$m and the radio continuum emission at 6\,cm from NGC~1365 before and after subtracting the central 80$\arcsec$ region. 
Cross-correlation coefficient $r_w(a)$ is larger on spatial scales $a$ where correlations are tighter.}
\label{fig:corr}
\end{figure}
\begin{figure*}
\begin{center}
\resizebox{14cm}{!}{\includegraphics*{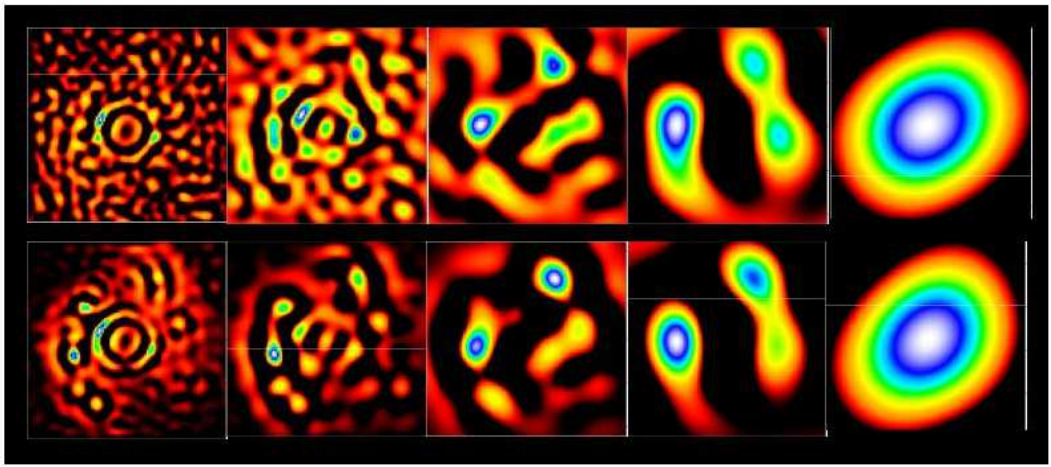}}
\caption[]{Wavelet decomposed maps at 870\,$\mu$m ({\it first row}) and 6\,cm ({\it second row}) on scales $a\simeq$\,2, 3.5, 6.4, 8, and 29\,kpc. }
\label{fig:slides}
\end{center}
\end{figure*}
\subsection{Origin of the observed emission}
Generally, the broad band emission at 870\,$\mu$m could consist of four main components: thermal
dust emission, free-free emission from thermal electrons, synchrotron
radiation from relativistic electrons, and contamination by CO(3-2) line emission. As we
are interested in the thermal dust emission alone, we
have to investigate the contribution of the other  components 
to the data.

The contribution of the CO(3-2) line emission to the surface brightness measured with the bolometer with a bandwidth of $\Delta \nu_{\rm bol}$ and a beam width of $\Omega_{\rm beam}$ can be calculated through 
\begin{equation}
F_{\rm line}= \frac{2 \kappa \nu^3 c^{-3}}{\Delta \nu_{\rm bol}} \Omega_{\rm beam} I_{\rm CO(3-2)},
\end{equation}
where $I_{\rm CO(3-2)}$ is the velocity integrated main-beam brightness temperature ($I_{\rm CO(3-2)}\,=\int T_{\rm mb}^{\rm CO(3-2)} . {\rm d}v)$) in K\,km\,s$^{-1}$.  Thus, $F_{\rm line}$\,[mJy]\,=\,0.973\,$I_{\rm CO(3-2)}$\,[K\,km\,s$^{-1}$] for the LABOCA  bandwidth of 60\,GHz and at 23\arcsec resolution. Using the SEST data, the contribution of the CO(3-2) line emission to the observed 870$\mu$m continuum emission varies in the range 16\%-25\% in different locations. In the central 80\arcsec area, the CO(3-2) flux is $\simeq$\,220\,mJy, i.e., 20\% of the observed 870$\mu$m flux ($S\,=\,1.1 \pm 0.2$\,Jy). The contribution of the CO(3-2) line emission was subtracted from the observed 870$\mu$m emission before studying the dust physical properties.  

In the core, where the contribution of the radio continuum emission has its maximum, the integrated flux density of the 6\,cm radio continuum emission is $\simeq$\,163\,mJy. The thermal free-free fraction at 6\,cm is about 20\% \citep{Beck_etal_05}.  Since the free-free flux changes with wavelength as $\lambda^{0.1}$, the corresponding thermal free-free flux at 870\,$\mu$m is  21\,mJy. Assuming a nonthermal spectral index of $\alpha_n=$0.8, the contribution of the synchrotron emission ($\sim \lambda^{\alpha_n}$) is about 4\,mJy. This is an upper limit, as the synchrotron spectrum is likely to steepen  due to CRE energy losses. Thus, only 1-2\% of the total 870\,$\mu$m flux is contaminated by the free-free and synchrotron emission.  

\subsection{Heating sources of cold dust in NGC~1365}
About 99\% of the energy released by galaxies in the FIR and submm wavebands is produced by thermal emission from dust grains. However, the energy sources which heat the dust and  power this emission  are often uncertain. Any effective source of optical/ultraviolet (UV) radiation, either young massive stars or an accretion disk surrounding an AGN, would heat dust grains.  Regions of intense dust emission are opaque at short wavelengths, and thus little information can be derived by optical or UV observations. As an extinction-free tracer of the ionized gas and star formation, the radio continuum emission can be used, instead, to probe the heating sources of dust.  Such studies are most informative when performed locally and at resolved scales in galaxies. Global studies are possibly biased toward the brightest emitting components in a galaxy. For example, the well-known radio-FIR correlation is weighted by regions of massive star formation when studied globally and in galaxy samples    (Tabatabaei et al. 2013). Only recently and through studying smaller scales within galaxies, variations of such a tight correlations have become apparent \citep[e.g. ][]{Hughes_etal_06, Tabatabaei_1_07,Tabatabaei_10,Dumas}.

We perform scale-by-scale comparison of the 870\,$\mu$m and the 6\,cm radio continuum emission  using a wavelet cross-correlation analysis. After convolution of the 6\,cm radio map to the resolution of the 870\,$\mu$m map (23\arcsec), the maps were normalized in grid size, reference coordinates, and field of view.  
The maps of the 870\,$\mu$m and 6\,cm emission were first decomposed into 10 scales from 23\arcsec ($\sim$2\,kpc) to about 300\arcsec ($\sim$27\,kpc) using the Pet-Hat wavelet function as detailed in \cite{Frick_etal_01}, \cite{Tabatabaei_1_07}, \cite{Laine}, and  \cite{Dumas}. 
Then, we cross-correlated the resulting decomposed maps of the 870\,$\mu$m and 6\,cm emission at each of the 10 spatial scales. In Fig.~\ref{fig:corr},  the cross-correlation coefficients $r_w$ (for pure correlation or anti-correlation $r_w$=$\pm$1) are plotted vs. the spatial scale $a$ before and after subtracting the central 80\arcsec.  Before subtracting the core, the two emissions are perfectly correlated as $r_w > 0.9$ on all scales. After the subtraction, however, the radio--submm correlation decreases particularly on scales $a<8$\,kpc. This shows that the good radio--submm correlation is mainly due to the core strongly emitting at both radio and submm  wavelengths, under starburst conditions. 

After subtracting the core, the situation resembles the radio--FIR correlation in normal star forming galaxies, where the correlation decreases toward small scales \citep[e.g. see][]{Hughes_etal_06,Dumas}. The decreasing trend of the radio--FIR correlation could be attributed to different origins of the radio continuum emission and the dust emission. For instance,  a weaker radio--FIR correlation is expected on small scales if the radio continuum emission is dominated by the synchrotron radiating cosmic ray electrons (CREs) diffused on large scales along the interstellar magnetic field lines \citep{Taba_13} or if the heating source of the dust is not linked to massive stars on small scales, but to a diffuse radiation field (ISRF). The latter is more likely the case for the cold dust emission traced at long FIR/submm wavelengths. 

Looking at the wavelet decomposed maps  (Fig.~\ref{fig:slides}), the greatest  difference in the morphologies is detected at the smallest scale ($\simeq2$\,kpc).  At this scale,  the radio emission exhibits few  point-like features as well as weaker filament-like structures following the spiral arms, while the cold dust emission shows dispersed clumpy structures. Such a non-coherent distribution is expected for diffuse emission which fits to the dust heating scenario by a diffuse ISRF. We also note that strong noise could also provide a non-coherent morphology on small scales reducing the correlation \citep{Dumas}. This, however, cannot be the entire reason of the observed  decreasing trend in the radio-FIR correlation in galaxies, as the decreasing trend resists using more sensitive Herschel data \citep{Taba_13}.  The morphologies of the radio and submm emission also differ significantly at the scale of 3.5\,kpc, becoming more similar toward larger scales.  At  $a=8$\,kpc, both radio and submm maps are similarly dominated by diffuse emission from star forming complexes in the ridges and along the spiral arms, leading to a perfect radio--submm correlation at this scale (see Fig.~\ref{fig:corr}).  
\begin{figure*}
\resizebox{\hsize}{!}{\includegraphics*{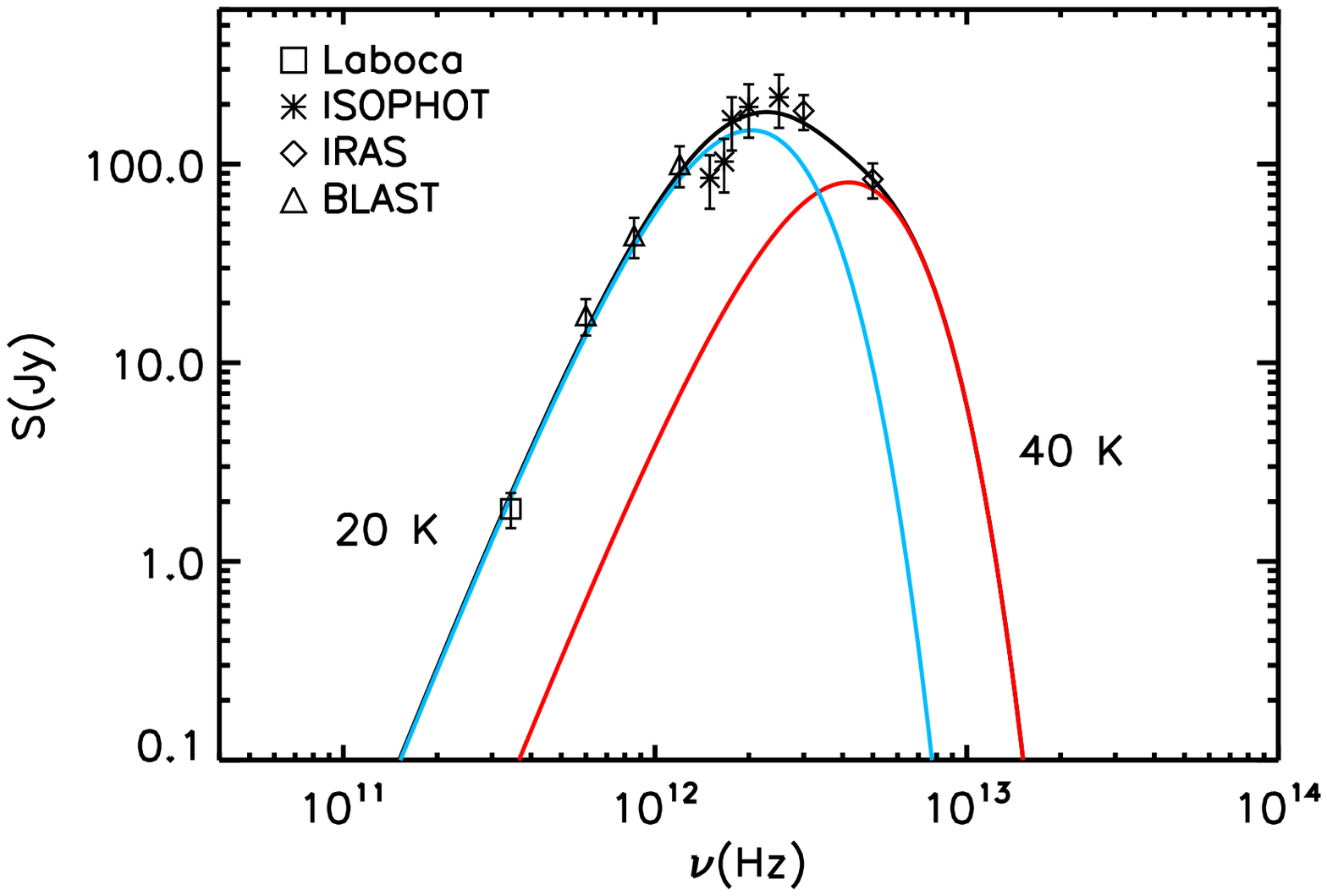}
\includegraphics*{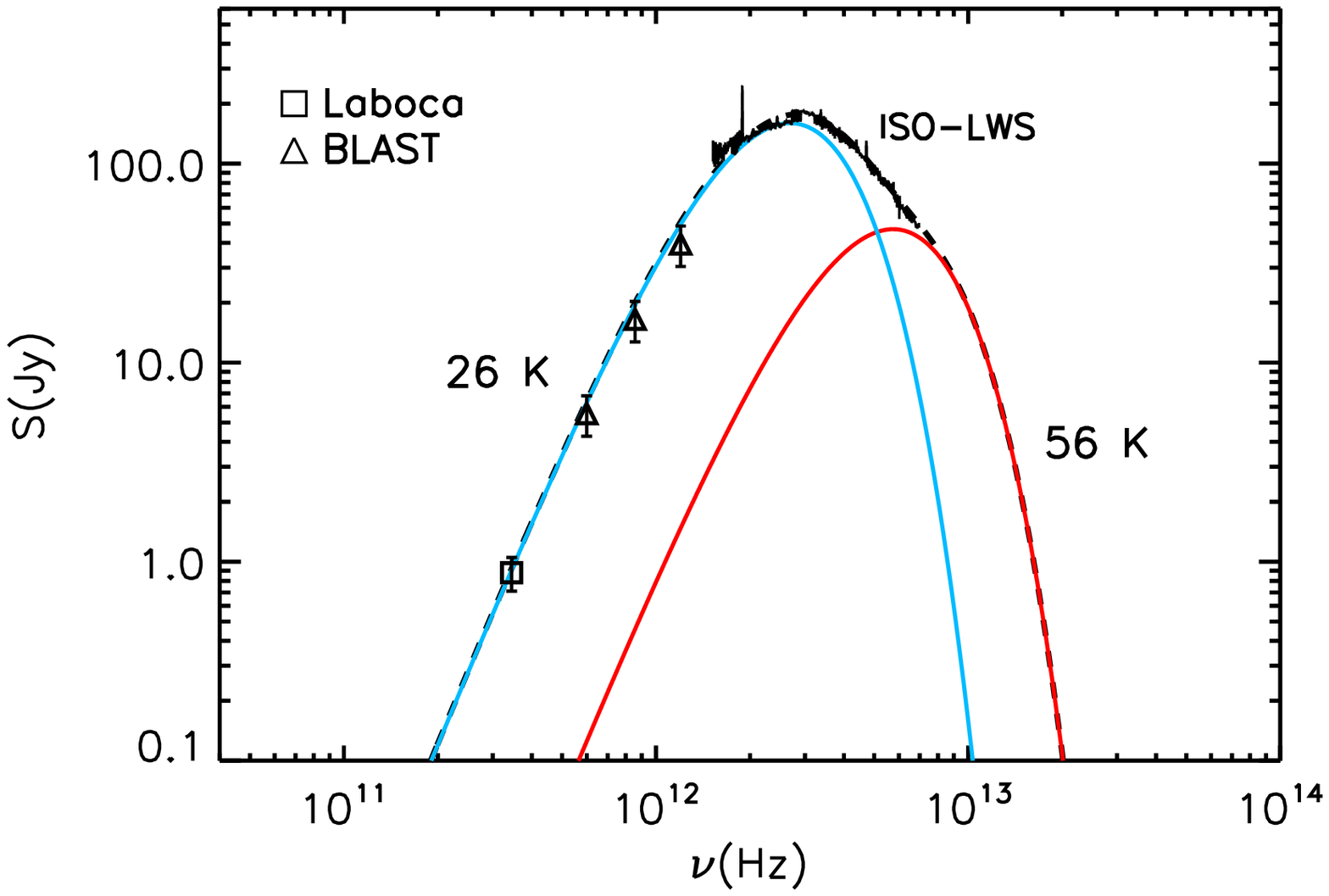}}
\caption[]{Spectral energy distribution of NGC~1365 ({\it left}) and of the central 80$\arcsec$ ({\it right}).  The colder component provides about 99\% of the total dust content in the central part, and 98\% in the entire area toward which 870\,$\mu$m emission is detected.  }
\label{fig:sed}
\end{figure*}   
\subsection{Dust physical properties}
We derive the dust mass and temperature assuming that dust grains are in local thermodynamic equilibrium (LTE) and hence  emit as a modified black body (MBB). Such a condition  applies for thermalized dust grains usually emitting at FIR and submm wavelengths (emission in the mid-IR, $\lambda<40\mu$m, is dominated by very small grains which are not thermalized).  
The dust SED derived based on the LABOCA and BLAST submm data together with the IRAS  and the ISOPHOT FIR data can be best re-produced if a two-component MBB is used (Fig.~\ref{fig:sed}): 
\begin{eqnarray}
S_{\nu} &=& \Omega_{\rm s}\, [B_{\nu}(T_{\rm c}) (1\,-\,e^{-\tau_{\nu,{\rm c} }}) \\
& &  + \, B_{\nu}(T_{\rm w}) (1\,-\,e^{-\tau_{\nu,{\rm w} }})] \nonumber,
\end{eqnarray}
where $S_{\nu}$ is the FIR/submm flux, $B_{\nu}$ the Planck function, $\nu$ the
frequency, and $\Omega_{\rm s}$ the solid angle of the emitting area subtended to the observer.  The two components, i.e., the cold and warm dust components are specified by their temperatures $T_{\rm c}$, $T_{\rm w}$ and mass surface densities $\Sigma_{\rm c}$, $\Sigma_{\rm w}$ given by their optical depths $\tau_{\nu, {\rm c}}$ and $\tau_{\nu, {\rm w}}$ as:
$$ \Sigma_{\rm c} = \tau_{\nu, {\rm c}}/ \kappa_{\nu},$$
$$ \Sigma_{\rm w}= \tau_{\nu, {\rm w}}/ \kappa_{\nu},$$
where $\kappa_{\nu}$ is the dust opacity or absorption coefficient. We adopt  $\kappa_{\nu} = 0.04\, (\frac{\nu}{250 {\rm GHz}})^{\beta}$ in units of m$^2$ per kilogram of a standard dust including silicates and amorphous carbon \citep[][chapter 14]{weiss_08,krugel}. 
Using a standard $\chi^2$ minimization technique,  the best fitted MBB model to the observed SED results in  a cold dust temperature $T_{\rm c}$ of 20\,K and a warm dust temperature  $T_{\rm w}$ of 40\,K. The mass surface densities are about 0.1\,M$_{\odot}$\,pc$^{-2}$ and 1.6$\times 10^{-3}$\,M$_{\odot}$\,pc$^{-2}$ for the cold and warm dust components, respectively. Thus,   about 98\% of the total dust content in this galaxy can be described with a temperature of 20\,K. This model is equally well described by dust emissivity indices in the range  $\beta=2.0 \pm 0.1$ (providing $\chi^2<2 \chi^2_{\rm min}$). 
This leads to a dust absorption coefficient at 870\,$\mu$m of $\kappa_{870}=0.076 \pm 0.002$\,m$^2$\,kg$^{-1}$ which is in agreement with \cite{James_02}. 

\begin{table*}
\begin{center}
\caption{Dust temperature and mass surface densities for disk and core of NGC\,1365.  }
\begin{tabular}{ l l l l l l l } 
\hline
Integrated region &  T$_{\rm c}$ (K)  &  T$_{\rm w}$ (K) &  $\Sigma_{\rm c}$ (M$_{\odot}$\,pc$^{-2}$) & $\Sigma_{\rm w}$ ($10^{-3}$\,M$_{\odot}$\,pc$^{-2}$) & M$_{\rm d}$ ($10^{7}$\,M$_{\odot}$)  \\
\hline 
\hline
  &   &   &  & & & \\
Disk   &   20\,$\pm$\,1 & 40\,$\pm$\,5 & 0.11\,$\pm$\,0.01 & $1.6\,\pm\,0.6$ & $10.1\pm0.8$ \\
 &   &   &  & & & \\
Core  & 26\,$\pm$\,1 & 56\,$\pm$\,4 & 0.64\,$\pm$\,0.06 & $4.3\pm1.3$ & $2.9\pm0.2$  \\
\hline
\end{tabular}
\tablefoot{The dust mass is also calculated for each area. The errors indicate the range of parameters which provide statistically good fits ($\chi^2<2 \chi^2_{\rm min}$).}
\end{center}
\end{table*}
We also derived the SED for the central 80$\arcsec$ area (core) for which the LABOCA submm data were used together with  the BLAST data and the ISO long wavelength spectrometer (LWS) data with a good coverage of the peak of the SED (Fig.~\ref{fig:sed}, b).  
In this region, the temperature of the  cold and warm dust components are  26\,K and 56\,K, respectively (see Table~3). The best fitted $\beta$ in the core is the same as in the disk.
The temperatures are in agreement with \cite{Alonso} who fitted the SED using the Herschel data.  
 
We derive a total dust mass of $M_{\rm d}\simeq 10^{8}\,{\rm M}_{\odot}$ for the entire galaxy taking into account only the points with intensities larger than 3\,$\sigma$ at 870\,$\mu$m.  This is in close agreement with \cite{Wiebe} taking into account the different absorption coefficients they used ($\kappa_{870}\simeq0.02$\,m$^2$\,kg$^{-1}$).
In the core, $M_{\rm d}\simeq 2.9 \times\,10^{7}\,{\rm M}_{\odot}$ constituting about 30\% of the total dust mass in the galaxy.

\begin{figure}
\begin{center}
\resizebox{6cm}{!}{\includegraphics*{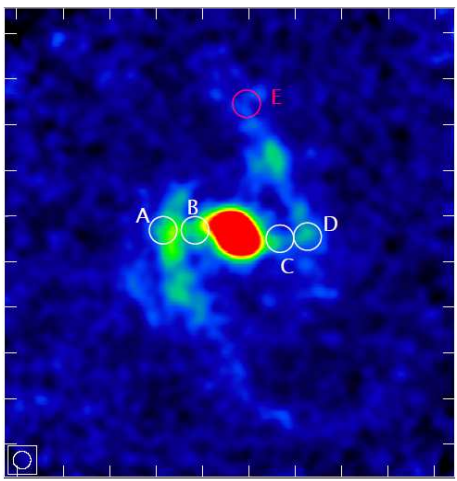}}
\caption[]{Selected apertures with 36\arcsec diameter along the bar and northern spiral arm superimposed on the 870\,$\mu$m map of NGC\,1365. }
\label{fig:apper}
\end{center}
\end{figure}

In a second approach, we investigate the dust physical properties along the bar using the BLAST 250\,$\mu$m-to-LABOCA 870\,$\mu$m flux ratios in apertures of 36$\arcsec$ (the angular resolution of the BLAST 250\,$\mu$m data). The color temperature can be derived using the following expression:
\begin{equation}
\frac{S_{250}}{S_{870}}=\frac{\nu_{250}^{\beta}}{\nu_{870}^{\beta}} . \frac{B_{250}(T)}{B_{870}(T)} \ , 
\end{equation} 
where $S_{250}$ and $S_{870}$ denote the measured flux at 250$\mu$m and 870$\mu$m, respectively. We use $\beta$=2 as derived based on the SED studies and that the dust is optically thin which is valid at the wavelengths considered. The corresponding dust mass is then given by 
\begin{equation}
M_{\rm d}=S_{870} D^2 \kappa_{870}^{-1} B_{870}(T)^{-1}
\end{equation}
Measurements of $S_{250}$ and $S_{870}$  in apertures shown in Fig.~\ref{fig:apper} lead to   $T$ and $M_{\rm d}$ values listed in Table~4. Along the bar, $T$ changes between 21$\pm$3\,K to 42$\pm$5\,K with errors determined based on $\sim$20\% calibration uncertainty of the integrated flux densities. The dust mass $M_{\rm d}$ changes between  $(0.8\pm 0.2)\times 10^{6}\,{\rm M}_{\odot}$  to $(6.8 \pm 1.8)\times 10^{6}\,{\rm M}_{\odot}$ in the selected apertures along the bar. The dust temperature in the apertures A \& B  (in the eastern bar) is similar to that of the spiral arms (see e.g.  aperture E) and also similar to that of the cold dust in the disk ($\simeq$20\,K) obtained based on the SED analysis. On the other hand, the dust is  warm in the western apertures C \& D (the average equilibrium temperature in these two apertures is $\simeq$\,40\,K the same as $T_{\rm w}$ for the disk). This together with the fact that the eastern bar is brighter than the western bar (Sect.~3.1) implies that the bar should contain more dust in the east than in the west. Table~4 shows that the east/west ratio  in $M_{\rm d}$ amounts to more than a factor of 4.

\begin{table*}
\begin{center}
\caption{Dust temperature and mass along the bar in apertures shown in Fig.~\ref{fig:apper}. }
\begin{tabular}{ l l l l l l} 
\hline
Aperture & RA & DEC &T$_{\rm c}$   &  M$_{\rm d}$ & M$_{\rm G}$  \\
 &[$^h$ $^m$ $^s$] &[$^{\circ}$ $\arcmin$ $\arcsec$] &[K] &[10$^6$\,M$_{\odot}$]  & [10$^8$\,M$_{\odot}$]\\
\hline 
\hline
A & 3 33 44.10 & -36 08 16.59 &21$\pm$3 & 3.2$\pm$0.8 & 4.9$\pm$0.8 \\
B & 3 33 40.59 & -36 08 23.16 &23$\pm$4 & 6.8$\pm$1.8 & 10.2$\pm$1.8\\
C & 3 33 31.67 & -36 08 39.54 &42$\pm$5 & 1.4$\pm$0.3 & 2.1$\pm$0.3\\
D & 3 33 27.75 & -36 08 29.70 &36$\pm$4 & 0.8$\pm$0.2 & 1.2$\pm$0.2\\
E & 3 33 36.53 & -36 05 21.40 &20$\pm$3 & 0.8$\pm$0.2 & 1.3$\pm$0.2\\
\hline
\end{tabular}
\end{center}
\label{tab:flux}
\end{table*}
\section{Discussion}   
\subsection{Molecular gas mass and X$_{CO}$ conversion factor}
The total dust mass determined from fitting to the SED of the disk ($R \leq 220\arcsec$) leads to a total gas mass $M_{\rm G}$  of $1.5 \times 10^{10}\,{\rm M}_{\odot}$ for a hydrogen gas-to-dust mass ratio of 150 \citep[see e.g.][and references therein]{krugel,Young_91}. The total gas surface density ($\sim$17\,M$_{\odot}\,{\rm pc}^{-2}$) is about 3 times larger than the local gas surface density in the Milky Way \citep[$\sim$ 6\,M$_{\odot}\,{\rm pc}^{-2}$][]{Dame_94}. 

In order to compare the gas mass estimate from the dust mass with that derived from the existing HI and CO observations the same integration area must be taken into account. Hence, we obtain the dust mass  for the same restricted area for which CO data are available \citep[204\arcsec$\times$164\arcsec,][]{Sandqvist_95}. The integrated 870\,$\mu$m flux, $S_{870}\simeq$\,1.52\,Jy, results in a dust mass of M$_{\rm d}\simeq 7.12\times10^7$\,M$_{\odot}$ using Eq.(4) for the 20\,K dust.  The corresponding total hydrogen gas mass is then M$_{\rm G}=1.07 \times 10^{10}\,{\rm M}_{\odot}$.

Integrating the HI map in the same area results in a  HI flux of $S_{\rm HI}=2.1\times10^4$\,Jy\,m\,s$^{-1}$. Using the calibration relation
$$ {\rm M}_{\rm HI}=2.356 \times 10^5 S_{\rm HI} D^2\,\,\,{\rm M}_{\odot},$$ 
with $S_{\rm HI}$  in Jy\,km\,s$^{-1}$ and  $D$ in Mpc \citep[e.g.][]{Jorsater_Moorsel}, we obtain  the HI mass of M$_{\rm HI}=1.72\times 10^9\,{\rm M}_{\odot}$. Thus, the mass of the molecular gas is M$_{\rm H_2}={\rm M}_{\rm G}\,-\,{\rm M}_{\rm HI}=\,8.95 \times 10^{9}\,{\rm M}_{\odot}$. This is about 48\% smaller (relative difference) than the H$_2$ mass estimate using the CO data and assuming a CO-to-H$_2$ conversion factor of $X_{\rm CO}=2.3\times 10^{20}$\,mol\,cm$^{-2}$\,(K\,km\,s$^{-1}$)$^{-1}$ \citep[M$_{\rm H_2}=1.73\times 10^{10}\,{\rm M}_{\odot}$,][]{Sandqvist_95}. The two M$_{\rm H_2}$ estimates would be  the same if a smaller  $X_{\rm CO}$ of $1.2\times~10^{20}$\,mol\,cm$^{-2}$\,(K\,km\,s$^{-1}$)$^{-1}$ is used. We stress that this value which is an upper limit, as derived using the lowest possible dust temperature and hence highest possible dust and gas mass,  is smaller than the default Galactic value of $2\times~10^{20}$\,mol\,cm$^{-2}$\,(K\,km\,s$^{-1}$)$^{-1}$.

For a similar comparison in the core, we first derive M$_{\rm H_2}$ using the CO data and assuming a CO-to-H$_2$ conversion factor of $X_{\rm CO}=2.3\times 10^{20}$\,mol\,cm$^{-2}$\,(K\,km\,s$^{-1}$)$^{-1}$ as used in \cite{Sandqvist_95}. The intensity $(I_{\rm CO}\,=\int T_{\rm mb} . {\rm d}v)$ of the CO(1-0) line averaged over the central 80\arcsec area,  
taking into account the SEST's beam width of 44$\arcsec$ \,at 110\,GHz is $\overline{I}_{\rm CO}$\,=\,41.5\,K\,km\,s$^{-1}$.   Following \cite{Sandqvist_95},  M$_{\rm H_2}=3.7\times 10^6\, L_{\rm CO}$ with the CO luminosity given by $L_{\rm CO}=~A\,\overline{I}_{\rm CO}$ ($A$ is the integrated area in kpc$^2$). Thus, the molecular gas mass in the bulge is M$_{\rm H_2}=~6.28\times 10^9\,{\rm M}_{\odot}$  using the CO data. On the other hand, based on the dust mass (see Table~3) and taking into account the HI mass, the molecular gas mass is M$_{\rm H_2}=~4.33\times\,10^{9}\,{\rm M}_{\odot}$ implying a $X_{\rm CO}$ conversion factor of $1.6\times~10^{20}$\,mol\,cm$^{-2}$\,(K\,km\,s$^{-1}$)$^{-1}$.
%

In the above estimate, it is assumed that the gas-to-dust ratio in the core is the same as for the  disk. However, the gas-to-dust ratio or metallicity usually shows a radial gradient in galaxies \citep[e.g.][]{Mateos,Tabatabaei_10,Magrini,Smith}. Such variations should be considered to estimate the $X_{\rm CO}$ conversion factor using the dust emission (FIR/submm) surveys   across a galaxy \citep[e.g.][]{Cox_86}.  Based on optical observations of 53 HII regions,  \cite{Pilyugin_04} obtained a radial gradient in metallicity or oxygen abundance  in NGC\,1365 as follows:
\begin{eqnarray}
Z & \equiv & \left. 12 + {\rm log}({\rm O/H})=\, \right.
\nonumber \\
& &\left. -(0.023\pm0.005)\,R({\rm kpc}) + (8.74\pm0.06)\right.
\end{eqnarray}
Generally, the relative amount of dust and gas is expected to 
be correlated with the abundance of the heavy elements \citep[see e.g.][]{Draine_2007}. 
Using a linear correlation between $Z$ and the dust-to-gas mass ratio $D$ \citep{James_02},  
Eq.~(5) leads to a gas-to-dust mass ratio of  $\simeq$117 in the core (assuming that it is 150 in the disk). This decreases the estimated molecular gas mass to M$_{H_2}=3.36~\times~10^9\,{\rm M}_{\odot}$ and  the conversion factor to $X_{\rm CO}=~1.2\times~10^{20}$\,mol\,cm$^{-2}$\,(K\,km\,s$^{-1}$)$^{-1}$ which is the same as in the central disk (the 204$\arcsec \times$\,164$\arcsec$ region).
 
On the other hand, it has been shown that $Z$ and $D$ could be correlated nonlinearly in galaxies \citep[e.g.][]{Issa,schmidt_93,Lisenfeld}.  The correlation given by \cite{schmidt_93} agrees with that given by \cite{Issa} (which includes few large galaxies like the Milky Way, M31, M51, and M101) but has a better statistics.  \cite{schmidt_93} found  that $Z$ is related to the logarithm of the dust-to-gas mass ratio $D$ through $Z\sim D^{0.63 \pm 0.25}$.  Assuming that the same proportionality applies in NGC\,1365,  we find  a gas-to-dust mass ratio of  $\simeq$100 in the core, leading to a molecular gas mass of M$_{H_2}=2.86~\times~10^9\,{\rm M}_{\odot}$ and a conversion factor of $X_{\rm CO}= 1.0\times 10^{20}$\,mol\,cm$^{-2}$\,(K\,km\,s$^{-1}$)$^{-1}$. Therefore, the $X_{\rm CO}$ conversion factor is smaller in the core than in the central disk by 20\%.  

Taking into account the metallicity gradient, the central gas mass concentration defined as the ratio of the total gas mass in the core to that in the entire disk is M$_{\rm core}$/M$_{\rm disk}\simeq$\,0.2.


%
%
%
\subsection{Star formation rate}
Integrating the modeled SED for $40\, \mu$m $ <\lambda< 1000\,\mu$m, the FIR luminosity is derived as L$_{\rm FIR}= 8.33 \times\,10^{10}\,{\rm L}_{\odot}$. Following the FIR definition by \cite{Rice}  the obtained luminosity in the range $42.5\, \mu$m $ <\lambda< 122.5\,\mu$m is L$_{42.5-122.5}=5.46 \times\,10^{10}\,{\rm L}_{\odot}$. This is in agreement with \cite{Rice} giving L$_{42.5-122.5}=5.49 \times\,10^{10}\,{\rm L}_{\odot}$ considering the different distance they used. Using the SED-based luminosity in the range $40\, \mu$m $ <\lambda< 500\,\mu$m, L$_{40-500}= 8.31 \times\,10^{10}\,{\rm L}_{\odot}$ together with 
the FIR to total infrared luminosity TIR ($8\, \mu$m $ <\lambda< 1000\,\mu$m) luminosity calibration given by  \cite{Chary} for a sample of Luminous InfraRed Galaxies (LIRGs) and starburst galaxies, we derive 
L$_{\rm TIR}=9.8 \times\,10^{10}\,{\rm L}_{\odot}$  \citep[in agreement with][giving L$_{\rm TIR}\simeq10^{11}\,{\rm L}_{\odot}$]{Sanders}.

Several authors have used the ${\rm L}_{\rm TIR}/{\rm L}_{\rm FUV}$ ratio to measure the extinction   \citep[e.g.][]{Calzetti_01,Verley_09,Montalto}. Here we calculate the extinction in the core and in the disk using this method.  \cite{Calzetti_01} and \cite{Calzetti_05}  found the following relation between the visual extinction and ${\rm L}_{\rm TIR}/{\rm L}_{\rm FUV}$ for starburst condition:
\begin{equation}
 {\rm A}_{\rm V} = {\rm C}\,\times\, 1.76\,\times\, {\rm log_{10}}\,(\frac{1}{1.68} \,\times\, \frac{\rm L_{\rm TIR}}{\rm L_{\rm FUV}} \,+\, 1), 
\end{equation}
With C=\,1 for emission from  diffuse ionized gas and C=\,0.44 for emission from stars.
Using the GALEX data, we derive L$_{\rm FUV}=1.02\times 10^9\,{\rm L}_{\odot}$ for the corresponding region in the disk. Assuming that the extinction is mainly caused for  emission from  stars, A$_{\rm V} \simeq$1.4 is obtained. This is equivalent to a FUV extinction  A$_{\rm FUV} \simeq$3.5, resulting in a de-reddened FUV luminosity of $${\rm L}^0_{\rm FUV}={\rm L}_{\rm FUV}\,e^{\rm A_{\rm FUV}/1.086} \simeq 2.59 \times 10^{10}\,{\rm L}_{\odot}.$$

The star formation rate based on the FUV emission is given by $${\rm SFR}_{\rm FUV} ({\rm M_{\odot}/yr}) = 1.40 \times 10^{-28}\,{\rm L}^0_{\rm FUV} $$ with ${\rm L}^0_{\rm FUV}$ in ergs$^{-1}$\,s$^{-1}$\,Hz$^{-1}$ \citep{Kennicutt_98}. This leads to ${\rm SFR}_{\rm FUV}\simeq$\,7\,M$_{\odot}$\,yr$^{-1}$ for the disk of NGC\,1365.
On the other hand, assuming that the energy source of the TIR emission is provided by  massive stars, the so-called `dust enshrouded star formation rate' can be derived using the TIR luminosity following \cite{Kennicutt_98},
$$ {\rm SFR}_{\rm TIR} ({\rm M_{\odot}/yr}) = 1.71 \times 10^{-10}\,{\rm L}_{\rm TIR} $$  with ${\rm L}_{\rm TIR}$ in L$_{\odot}$. The corresponding value for NGC\,1365 is ${\rm SFR}_{\rm TIR}\simeq$\,16.7\,M$_{\odot}$\,yr$^{-1}$.

Table~5 shows similar calculations for the core (central 80\arcsec). The extinction value is in agreement with \cite{Kristen} who derived A$_{\rm V}\sim 2-2.5$ by means of the Balmer-decrement-ratio method.

The molecular depletion time scale, defined as the molecular gas mass per star formation rate (M$_{\rm H_2}$/SFR$_{\rm FUV}$), is about 1.2\,Gyr in the central disk (the $204\arcsec\times 164\arcsec$ area). However, in the core, it is $\simeq$\,0.7\,Gyr and 0.9\,Gyr for the nonlinear and linear $Z$--$D$ correlations, respectively.  This is due to a more efficient star formation  in the core than in the disk in NGC\,1365.

\begin{table}
\begin{center}
\caption{Properties of the disk and core in NGC\,1365. }
\begin{tabular}{ l l l l} 
\hline
Parameter & Unit &  Disk  &  Core  \\
\hline 
\hline
L$_{\rm TIR}$ & $[10^{10}$L$_{\odot}]$  & 9.8    & 8.7 \\
L$^0_{\rm FUV}$ & $[10^{10}$L$_{\odot}]$ & 2.5   & 1.4\\
M$_{\rm G}^{1}$     & $[10^{9}$M$_{\odot}]$   & 15.1   & 2.9 (3.4)$^{2}$\\
$X_{\rm CO}$ & $X_0^{3}$               &  1.2$^{4}$& 1.0 (1.2)$^{2}$\\
A$_V$           & $[$mag$]$    & 1.4  & 2.1  \\
A$_{\rm FUV}$  &   $[$mag$]$  & 3.5    & 5.3 \\
SFR$_{\rm TIR}$ &$[$M$_{\odot}$yr$^{-1}]$   & 16.7& 15.0\\
SFR$_{\rm FUV}$ & $[$M$_{\odot}$yr$^{-1}]$ & 7.0  & 3.9\\
\hline
\noalign {\medskip}
\multicolumn{4}{l}{$^{1}$ The total gas mass}\\
\multicolumn{4}{l}{$^{2}$ For a nonlinear (linear) $Z$-$D$ correlation}\\
\multicolumn{4}{l}{$^{3}$ $X_0=10^{20}$\,mol\,cm$^{-2}$\,(K\,km\,s$^{-1}$)$^{-1}$}\\
\multicolumn{4}{l}{$^{4}$ Calculated for {the central} $204\arcsec\times 164\arcsec$ area in the disk }\\
\end{tabular}
\end{center}
\end{table}

\subsection{Gas flow in the bar}
 \label{torq}
The central gas mass concentration obtained in Sect.~4.1 is evolving fast due to the gas flow in the bar.
The correlation between the cold dust emission and a sum of the atomic and molecular gas emission (Sect.~3.1) already shows that the cold dust is a proxy for the gas in the disk. This further motivates us to investigate the gas flow in the bar based on the sub-mm data as a tracer of the total gas. The strongly barred galaxy, NGC\,1365, is expected to experience gravity torques exerted
by the bar on the gas disk  which could efficiently drive the gas towards the center
through a reduction in its angular momentum.  

\begin{center}
\begin{figure}[ht]
\includegraphics[angle=-90,width=7.5cm]{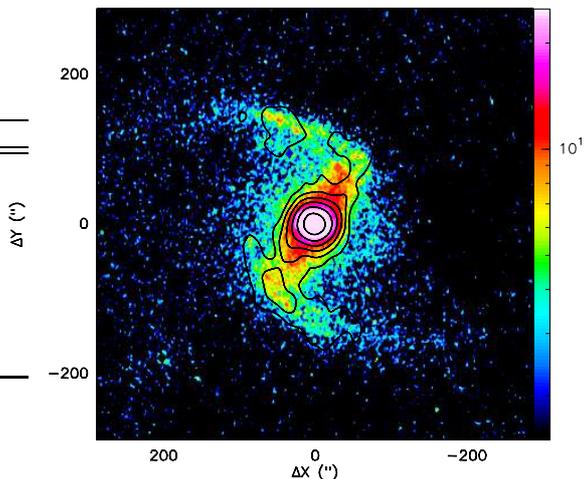}
\caption{ Logarithmic contours of dust emission 
superposed on  the near-infrared H image,
from 2MASS in logarithmic levels.
Both images have been deprojected, and rotated (50$^{\circ}$ counterclockwise) such that the major axis
is horizontal.}
\label{Hdust}
\end{figure}
\end{center}

We quantify the gas inflow in NGC\,1365 by averaging the action of gravitational forces on the gas at different radii following \cite{com81} and e.g. \cite{Garcia}.  The gravitational forces are computed to derive the underlying gravitational potential. It is assumed that the total mass budget is dominated by the stellar contribution and that the effect of gas self-gravity can be neglected.
As a proxy of the stellar  mass distribution, we use the H-band (1.5\,$\mu$m) image of the 2MASS data,  being only weakly affected by dust extinction or by stellar population biases. After removing the foreground stars, the H-band image is deprojected (the position angle and inclination are listed in Table~1). The image was then resampled at 1.5'' per pixel.
The deprojected H-band image is superposed on the dust contours
in Fig.~ \ref{Hdust}, showing a very good correspondence. This indicates that 
the molecular gas is well aligned along the bar
and spiral arms in NGC\,1365. However, a slight phase shift can be
noticed. The dust is shifted to the leading side of the bar: in Fig.~8, the
contours of the dust in the bar are lemon-shape elongated ellipsoids,
which extremities are shifted to smaller position angles with respect to the red-color
bar on both sides, i.e. north and south of the center.

\begin{figure}[ht]
\includegraphics[angle=-90,width=7cm]{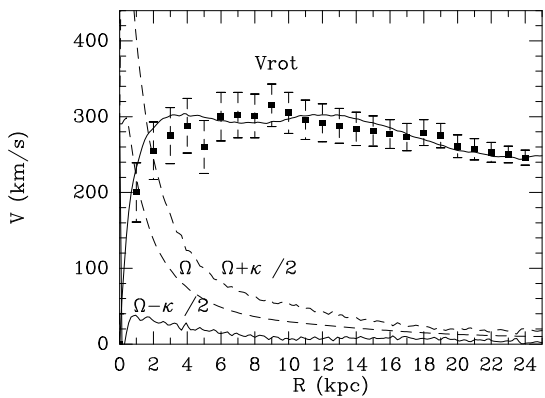}
\caption{ Rotation curve and derived frequencies $\Omega$,
$\Omega-\kappa/2$ and $\Omega+\kappa/2$, for NGC 1365, obtained from the H-band
image, and a constant M/L ratio. The model rotation curve has been fitted to 
the data points compiled from H$\alpha$ and HI data  (see Zanmar Sanchez et al 2008). } 
\label{vcir}
\end{figure}

The deprojected H-band image is  completed in the vertical
dimension by assuming an isothermal plane model with a constant scale height,  
equal to $\sim$1/12th of the radial scale-length of the image. The potential is
then derived by a Fourier transform method, assuming a constant mass-to-light (M/L) ratio. 
The M/L value is selected to retrieve  the observed rotation curve \citep[given by][using H$\alpha$ and HI data]{zanmar}.
Only a very light dark matter halo is added, of 3$\times$10$^{10}$ M$_\odot$, to better fit
the rotation curve in the outer parts.
The axisymmetric part of the model, fitted by parametric functions, is then derived to find
the proper frequencies, as shown in Fig.~ \ref{vcir}.

For the non-axisymmetric part, the
potential $\Phi(R,\theta)$  is then decomposed in the different m-modes:
$$
\Phi(R,\theta) = \Phi_0(R) + \sum_m \Phi_m(R) \cos (m \theta - \phi_m(R))
$$
\noindent
where $\Phi_m(R)$ and $\phi_m(R)$ represent the amplitude and phase of the m-mode.

\begin{figure}[ht]
\includegraphics[angle=-90,width=7cm]{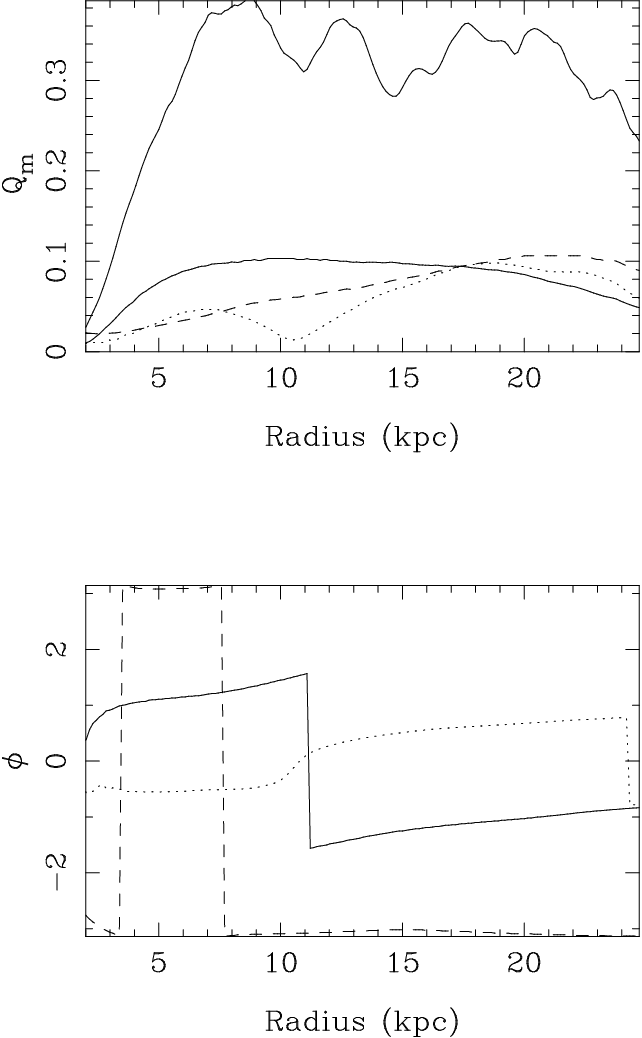}
\caption{Strengths (Q$_1$, Q$_2$,  Q$_4$ and total Q$_T$) and phases ($\phi_1$,  $\phi_2$ and $\phi_4$) of the $m=1$ (dash) $m=2$ (full line) and $m=4$ (dots) Fourier
components of the stellar potential. The region of constant $m=2$ phase delineates the extent of the bar (note the phase jumps by 2$\pi$/$m$). The $\phi$-angles
are measured from the +X axis in the counter-clockwise direction.}
\label{pot1365}
\end{figure}

Following \cite{com81}, we define the strength of the $m$-Fourier component, $Q_m(R)$,  as
$Q_m(R)={m \Phi_m \over R | F_0(R) |}$, i.e. by the ratio between tangential
and radial forces.
The strength of the total non-axisymmetric perturbation is defined by:
$$
Q_T(R) = {F_T^{max}(R) \over F_0(R)} 
$$
\noindent
where $F_T^{max}(R)$ represents the maximum amplitude of the tangential force and $F_{0}(R)$ is the mean axisymmetric radial force. This quantity is a measure of the strength of the bar. The variation of phase $\phi_m$ with radius $R$ discriminates between bar and spiral arms. For example, the phase is constant for m=2 in the bar-like potential, hence the extent of the bar can be deduced where $\phi_2(R)$=constant (see Figs.~\ref{pot1365}).  
A main bar can be seen clearly, together with two spiral arms, with small pitch angle.

After having calculated the 2D force field per unit mass ($F_x$ and $F_y$) from the derivatives of $\Phi(R,\theta)$ on each pixel, the torques per unit mass are derived ($t$(x, y) = x Fy - y Fx).
This torque field, by definition, is independent of the present gas distribution in the plane. 

\begin{figure}[ht]
\includegraphics[angle=-90,width=7cm]{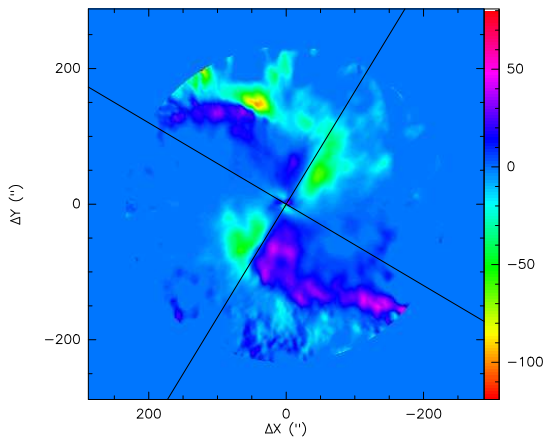}
\caption{Map of the gravitational torque 
(t(x,y)~$\times$~$\Sigma$(x,y), as defined in text) in the center of NGC~1365. 
The derived torques change sign as expected, following a pattern of
four quadrants. The orientation of quadrants follow
the bar orientation in NGC\,1365. In this deprojected picture,
the major axis of the galaxy is oriented parallel to the horizontal axis.}
\label{torq1365}
\end{figure}

The next steps consist of using the torque field to derive the angular
momentum variations and the associated flow time-scales.
We assume that the cold dust emission at each offset
in the galaxy plane is a fair estimate of the probability of finding
gas at this location at present. Hence, the gravitational torque map weighted by the gas surface density traced by the cold dust emission ($t(x,y)\times \Sigma(x,y)$, see Fig.~\ref{torq1365}) allows us to derive the net effect on the gas, at each radius (the torque map is oriented according to the sense of rotation in the galactic plane).

To estimate the radial gas flow induced by the torques, we first computed the torque 
per unit mass averaged over the azimuth, using $\Sigma(x,y)$ as the actual weighting function,i.e.:
$$
t(R) = \frac{\int_\theta \Sigma(x,y)\times(x~F_y -y~F_x)}{\int_\theta \Sigma(x,y)}
$$

\begin{figure}[ht]
 \includegraphics[angle=-90,width=7cm]{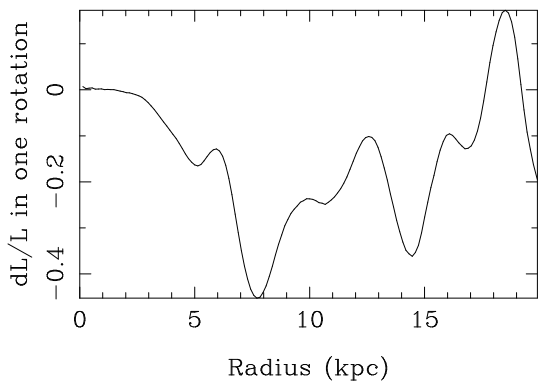}  
    \caption{Radial variation of the torque, or more precisely the fraction of the angular momentum transferred
from/to the gas in one rotation.}
   \label{gastor1365}
 \end{figure}

By definition, $t(R)$ represents the time derivative of the specific angular momentum $L$ of the gas averaged azimuthally, i.e., $t(R)$=$dL/dt~\vert_\theta$. To derive non-dimensional  quantities, we normalized this variation of angular momentum per unit time,
to the angular momentum at this radius, and to the rotation period.
We then estimate the efficiency of the gas flow with
the average fraction of the gas specific angular momentum transferred in one rotation 
(T$_{rot}$) by the stellar potential, as a function of radius, i.e., by the function $\Delta L/L$ defined as:
$$
{\Delta L\over L}=\left.{dL\over dt}~\right\vert_\theta\times \left.{1\over L}~\right\vert_\theta\times 
T_{rot}={t(R)\over L_\theta}\times T_{rot}
$$
\noindent
where $L_\theta$ is assumed to be well represented by its axisymmetric estimate, i.e.,$L_\theta=R\times v_{rot}$. 
 The  $\Delta L/L$ radial distribution for NGC\,1365 derived from the dust emission  is displayed in
Figs.~\ref{gastor1365}.

Fig.~\ref{torq1365} shows that the derived torques change sign following a characteristic
four quadrant pattern. There is only a notable exception in the top 
quadrant of the diagram, where a patch of strong negative torque
exists in the positive torque quadrant.  These perturbations could be the consequence
of infalling material, as noticed by \cite{zanmar}.
The gas location is however
mainly concentrated in the negative torque regions, as can be
seen by comparison with Fig.~ \ref{Hdust}, i.e.  
the majority of the gas in the bar is phase-shifted towards the leading edge, where the torques
are negative. The rotation sense in the galaxy  is clockwise, and the spiral
structure is trailing.

These results can be explained according to the classical
scenario of angular momentum transfer in barred galaxies.
The main bar of $\sim$ 11 kpc in radius ends slightly inside its co-rotation. 
This length obtained is larger than measurements using the I-band image of the galaxy \citep[100$\arcsec \sim$9\,kpc][]{zanmar}. 
From Fig. ~\ref{vcir}, it is deduced that the bar pattern speed is $\sim$ 26 km/s/kpc.
This value of the pattern speed indicates that 
there must be an inner Lindblad resonance (ILR) in the circum-nuclear region,
which might correspond to the large gas concentration there,
at r$<$ 2 kpc. In the central region, the stellar kinematics
suggest the presence of a nuclear disk, as shown by \cite{Emsellem}.
  
The gas flow towards the center is taken just in the act in NGC\,1365 at the present epoch.
Fig. ~\ref{gastor1365} shows that at 7 kpc radius, about 45\%  of the angular momentum
is removed in one orbit, so the radial flow time-scale is 300~Myr. Hence, the flow rate is relatively high due to the strong bar.
This leads to a large concentration of molecular gas towards the center, which must be
recent, since the center does not host a strong starburst currently.

These results indicate that the galaxy's AGN (Seyfert 1.5) is
fueled by the present gas flow.  However, it is not possible 
to conclude unambiguously whether the gas is stalled at the inner Lindblad resonance of the 
main bar or is still driven inwards due to a nuclear bar,
because of lack of spatial resolution.

There is a good correspondence between the regions of strongest (positive) torques (Fig.~11) and where ordered magnetic fields are strongest in NGC\,1365 (Beck et
al.  2005). This further shows that ordered magnetic
fields (traced by polarized radio emission) are signatures of
non-axisymmetric gas flows and hence angular momentum transfer. 

\section{Summary}
We produced the first large scale map of the giant barred galaxy NGC\,1365  at 870$\mu$m using the Large APEX Bolometer Camera at 20\arcsec \,resolution. The sub-mm map exhibits strong emission from the core and the bar, similar to  molecular gas traced by CO emission, as well as the large scale emission from the spiral arms, similar to HI emission. We investigate possible origins of this emission and perform dust SED analysis leading to estimates of the dust mass and total infrared luminosity.
Assuming that the cold dust, presented by the submm emission, traces the total neutral gas  in the galaxy, we further estimate the gas mass, the $X_{\rm CO}$ conversion factor (taking into account the variation in metallicity), and the star formation rate in the disk and the core (central 80$\arcsec$) of NGC\,1365.  
The most important findings of this study are summarized as follows:

\begin{itemize}
\item[-] The thermalized dust SED in NGC\,1365 can be best fitted by a 2-component modified black body model, with temperatures of 20\,K and 40\,K for cold and warm dust, respectively. The cold dust represents about 98\% of the total dust content in this galaxy.

\item[-] Comparing the gas mass obtained from the dust mass measurements with that based on the CO and HI observations, we derive an average CO-to-H$_2$ conversion factor of $X_{\rm CO}\simeq~1.2\times 10^{20}$\,mol\,cm$^{-2}$\,(K\,km\,s$^{-1}$)$^{-1}$ for the central disk (limited in a $204\arcsec\times 164\arcsec$ area).  This value is the same (20\% larger than) in the core, taking into account the metallicity variation and assuming a linear (nonlinear) correlation between the gas-to-dust mass ratio and the metallicity.  

\item[-] The central gas mass concentration reduces from $\sim$30\% to about 20\% taking into account  metallicity variations. 

\item[-] Integrating the dust SED, the total IR luminosity is L$_{\rm TIR}=9.8 \times\,10^{10}\,{\rm L}_{\odot}$ leading to a dust-enshrouded star formation rate of ${\rm SFR}_{\rm TIR}\simeq$\,16.7\,M$_{\odot}$\,yr$^{-1}$ in NGC\,1365. The star formation efficiency is found to be larger in the core than in the disk by $\gtrsim$50\%.
 
\item[-] The bar exhibits an east-west asymmetry in the 870\,$\mu$m emission similar to that in the 6\,cm radio continuum emission. This further leads to  an asymmetry in the distributions of the dust properties: The eastern bar is colder and heavier than the western bar by more than a factor of 4.  

\item[-] Apart from the similar distribution of the radio and submm emission along the bar and spiral arms, their correlation decreases by decreasing the spatial scale. This could indicate different origins of the cold dust emission (e.g.  heating by a diffuse ISRF) and the radio continuum emission (e.g. CREs propagated along the magnetic fields) rather than massive star formation.   

\item[-] Based on the cold dust map, it is deduced that the gas in NGC\,1365 flows towards the center with a time-scale of 300\,Myr. About 45\% of the angular momentum is removed in one orbit at 7\,kpc radius.

\end{itemize}
\acknowledgement We are grateful to Aa. Sandqvist for kindly providing us with the CO(3-2) data. We thank A. Belloche for useful discussion on LABOCA data reduction. FST acknowledges the support by the DFG via the grant TA 801/1-1.

\bibliography{s.bib}

\end{document}

%% file: thesiscom.tex
\DeclareRobustCommand{\ion}[2]{%
\relax\ifmmode
\ifx\testbx\f@series
{\mathbf{#1\,\mathsc{#2}}}\else
{\mathrm{#1\,\mathsc{#2}}}\fi
\else\textup{#1\,{\mdseries\textsc{#2}}}%
\fi}

\def\apjl{ApJ}%
\def\apjs{ApJS}%
\def\ao{Appl.~Opt.}%
\def\aap{A\&A}%
